\documentclass[preprint,12pt,authoryear]{elsarticle}

\usepackage[utf8]{inputenc}
\usepackage{amsmath}
\usepackage{graphicx}
\usepackage{hyperref}
\usepackage{booktabs}
\usepackage{float}
\usepackage{listings}
\usepackage{tikz}
\usepackage{array}

\hypersetup{
  colorlinks=true,
  linkcolor=blue,
  citecolor=blue,
  urlcolor=cyan
}

\journal{Transportation Research Part B: Methodological}

\begin{document}

\begin{frontmatter}

\title{Recurring Public Transit Schedules: Stable Identification from GTFS and Similarity Analysis}

\author[swat]{Evgeny Makarov\corref{cor1}}
\ead{evgeni@swatmobility.com}

\author[wsp]{Georgy Taubkin}
\ead{gera.taubkin@wsp.com}

\cortext[cor1]{Corresponding author}

\affiliation[swat]{organization={SWAT Mobility PTE LTD},%
  country={Singapore}}

\affiliation[wsp]{organization={WSP},%
  country={Canada}}

\begin{abstract}
Public transit schedules contain recurring service structures: many dates share the same passenger-facing timetable, while others differ because of weekends, holidays, seasons, or special events. GTFS does not encode these recurring schedules directly, so downstream scheduling, assignment, and mismatch models lack an explicit recurrent supply object. This paper formalizes recurring schedules as \emph{DayTypes}, defined by the complete set of Route Pattern trips operating on a date, and develops a stable GTFS-based extraction method using H3 route-pattern keys. We then define a schedule-comparison framework with exact, time-tolerant, and structural-comparability metrics that distinguish strict timetable differences from small timing shifts and larger service changes. Validation on Japanese and Canadian GTFS feeds shows substantial schedule compression, a median of four nonempty DayTypes per agency in the pairwise-analysis sample, hierarchical nesting between service classes, and country-level differences in the persistence of exact disjointness. The resulting DayTypes provide compact Route-Pattern-time schedule sets for timetable synchronization, vehicle scheduling, demand assignment, and schedule-consolidation analysis.
\end{abstract}


\end{frontmatter}

\section{Introduction}

Public transit planning requires a compact and identifiable representation of recurring service. Timetables are not independent from date to date: many calendar dates share the same passenger-facing supply, while others differ because of weekends, holidays, school terms, seasons, or special events. Planners and modelers must recover these recurring structures before they can compare service levels, design timetables, allocate vehicles, diagnose mismatch, or evaluate changes across agencies and feed versions.

A foundational concept for this temporal structure is the \emph{DayType}: a class of calendar dates that share operationally equivalent service. Transit agencies routinely use day types in practice---distinguishing weekdays, Saturdays, Sundays, holidays, and seasonal service periods. The Chicago Transit Authority, for example, publishes ridership statistics by weekday, Saturday, and Sunday/holiday categories \citep{cta2026}. Yet DayType is rarely formalized as a mathematical object, and it is not explicitly represented in standard transit data formats.

Transit data are now widely distributed through the General Transit Feed Specification (GTFS), the de facto global standard for public transit schedules \citep{mchugh2013}. GTFS has enabled extensive research on transit networks, including spatial analysis \citep{huynh2018}, spatio-temporal visualization \citep{prommaharaj2020}, and accessibility measurement \citep{liu2025}. However, GTFS encodes service through interdependent calendar, trip, stop-time, stop, and shape tables rather than through explicit recurring schedule objects. Thus a basic planning question---which recurring timetable operates on this date?---becomes a reconstruction and identification problem.

This paper treats DayType identification as a methodological representation problem upstream of timetable optimization, vehicle scheduling, network design, demand assignment, and activity-schedule analysis. These models require recurring supply objects: service classes, Route Patterns, and route-pattern-time tables against which synchronization, vehicle allocation, assignment, mismatch, or consolidation can be evaluated. The contribution is not a new timetable optimizer; it is the identifiable supply representation and comparison layer needed before such models can be applied consistently to GTFS-scale schedule data. We make this representation identifiable from GTFS, define route-pattern identifiers that are stable under administrative GTFS identifier relabeling, and compare DayTypes through a hierarchy of exact, time-tolerant, and structural-comparability metrics.

The framework is organized around three methodological requirements. First, the extracted DayType partition should be identifiable from a fixed feed and invariant to administrative relabeling of GTFS identifiers. Second, schedule comparison should remain interpretable under small timetable perturbations: one-minute shifts should remain visible for audit, but should not automatically imply structural service redesign. Third, the extracted objects should be usable as recurring schedule sets for downstream models that operate on timetables, Route Patterns, vehicle schedules, demand assignment, or activity schedules.

\subsection{Contributions}

This paper makes the following contributions:

\begin{enumerate}
  \item We formalize Route Patterns and DayTypes as recurring public-transit supply objects. A Route Pattern is a stable passenger-facing service variant, and a DayType is the complete set of Route Pattern trips operating on a calendar date.
  \item We develop a GTFS-based extraction procedure that reconstructs these objects from calendar, trip, stop-time, stop, and shape information. H3 route-pattern keys reduce dependence on unstable GTFS identifiers and give deterministic identifiability for a fixed feed.
  \item We define a schedule-comparison hierarchy: exact metrics for strict timetable audit, time-tolerant matching for bounded timing shifts, and structural-comparability diagnostics for separating timing-sensitive differences from trip insertion, removal, and Route Pattern change.
  \item We connect the extracted DayType object to downstream methodological models by treating it as a recurring Route-Pattern-time schedule set for synchronization, vehicle scheduling, demand assignment, mismatch diagnosis, and schedule consolidation.
  \item We validate the framework on Japanese and Canadian GTFS feeds, showing that recurring service structures can be extracted at scale and that the metric hierarchy changes the interpretation of apparently disjoint schedules.
\end{enumerate}

\section{The Transport Planning Perspective: Why DayTypes Matter}
\label{sec:planning_perspective}

Transit planning routinely groups dates with similar service requirements. This section positions DayTypes in that planning hierarchy before formalizing them mathematically.

\subsection{DayTypes in the Planning Hierarchy}

In the classical planning framework, network design and frequency setting precede timetable development, while vehicle and crew scheduling are constructed after a timetable has been specified \citep{ceder1986}. DayTypes enter this hierarchy as date-level timetable classes: they identify which recurring passenger-facing timetable is used before vehicle blocks, crew duties, and depot assignments are determined. Instead of treating every calendar date as a separate timetable, DayTypes reduce the calendar to recurring service classes such as weekdays, Saturdays, Sundays, holidays, school terms, seasons, or special-event days when these imply different service.

Timetable synchronization and demand-driven timetable design require service categories with common demand and operating conditions \citep{ceder2001,ceder2015}. A DayType is therefore a recurring timetable class, not merely a calendar label. It is defined by passenger-facing service---which trips operate, on which Route Patterns, and at what times---rather than by vehicle blocks, crew duties, depot assignments, fleet allocation, or maintenance plans. This distinction is important because DayType extraction from GTFS recovers the published supply pattern, not the internal operating plan.

\section{Problem Statement and Formal Definitions}
\label{sec:definitions}

We now define the objects needed to identify and compare DayTypes. The central task is to reconstruct, for each calendar date, the complete passenger-facing timetable as a set of Route Pattern trips.

\subsection{Operational Definitions}

We use four operational concepts: route, Route Pattern, trip, and DayType.

\subsubsection{Route (Passenger Route)}

A \textbf{route} is the passenger-facing service label, such as a route number or name. It is an organizational object: one route may contain multiple directions, deviations, short turns, or express variants.

\subsubsection{Route Pattern (Route Variant)}

A \textbf{Route Pattern} is a passenger-facing variant of a route, defined by an ordered stop sequence and travel direction, optionally associated with a physical path. Route Patterns distinguish inbound and outbound service, short turns, extensions, deviations, express variants, and loops. GTFS does not provide stable Route Pattern objects; they must be inferred from trips, stop times, stops, and shapes.

\subsubsection{Trip (Service Trip)}

\begin{figure}[H]
  \centering
  \includegraphics[width=\textwidth]{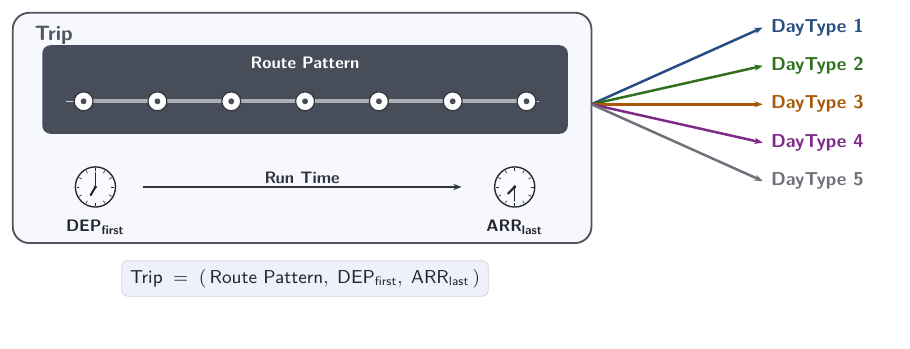}
  \caption{Conceptual diagram of a Trip as a tuple of Route Pattern, departure time from the first stop ($\text{DEP}_{\text{first}}$), and arrival time at the last stop ($\text{ARR}_{\text{last}}$). A single trip specification may appear across multiple DayTypes.}
  \label{fig:trip-diagram}
\end{figure}

A \textbf{trip} is one scheduled traversal of a Route Pattern. The trip specification used in this paper consists of the Route Pattern, first departure time, and last arrival time (Figure~\ref{fig:trip-diagram}). The same trip specification may appear in multiple DayTypes.

\subsubsection{DayType (Operational Definition)}

A \textbf{DayType} is a category of dates that share the same passenger-facing schedule: the same Route Pattern trips at the same times. It is a timetable object, not a vehicle-blocking, crew-scheduling, depot, fleet, or maintenance object. Vehicle blocks and crew duties are solved given a timetable and may vary even when the DayType is unchanged \citep{david2020}. This passenger-facing definition allows comparisons across agencies, feed versions, and planning approaches without requiring internal operating data.

\subsection{Mathematical Formalizations}

The operational definitions above lead to the following mathematical objects.

\subsubsection{Route Pattern (Mathematical Definition)}

Formally, let:

\begin{itemize}
  \item $R$ be the set of all routes in the transit network;
  \item $P_r$ be the set of all Route Patterns for route $r \in R$;
  \item $S_p = (s_1, s_2, \ldots, s_n)$ be the ordered stop sequence of Route Pattern $p \in P_r$;
  \item each stop $s_i$ have geographic coordinates $(lat_i, lon_i)$.
\end{itemize}

A Route Pattern $p$ is uniquely characterized by:

\begin{equation}
p = (S_p, \text{geography})
\end{equation}

where $S_p$ is the ordered stop sequence and $\text{geography}$ represents the physical path between stops. Direction of travel is encoded by the stop order: a northbound pattern and its corresponding southbound pattern have reversed stop sequences and are therefore distinct Route Patterns.

\subsubsection{DayType (Mathematical Definition)}

For date $d$, let $T_{p,d}$ be the set of trips operating on Route Pattern $p$. Each trip $t \in T_{p,d}$ has first departure time $dep_t$ and last arrival time $arr_t$. The atomic \textbf{trip specification} is:

\begin{equation}
x_t = (p, dep_t, arr_t).
\end{equation}

The \textbf{DayType} for date $d$ is the set of all such trip specifications:

\begin{equation}
\text{DayType}(d) = \{(p, dep_t, arr_t): r \in R, p \in P_r, t \in T_{p,d}\}.
\end{equation}

This set is the published passenger-facing supply for date $d$: every Route Pattern trip and its first departure and last arrival time.

Two calendar dates $d_1$ and $d_2$ belong to the \textbf{same DayType} (denoted $d_1 \equiv d_2$) if and only if:

\begin{equation}
\text{DayType}(d_1) = \text{DayType}(d_2)
\end{equation}

This exact definition intentionally treats any Route Pattern, departure-time, or arrival-time difference as a DayType difference. Section~\ref{sec:analysis-methods} then introduces tolerant comparison metrics that test whether exact differences persist after small same-Route-Pattern timing shifts.

\subsection{Extraction from GTFS}

With the formal objects established, we now describe how Route Patterns and DayTypes are extracted from standard GTFS feeds.

\subsubsection{Route Pattern Extraction from GTFS}

For each trip, we extract the ordered stop sequence from \texttt{stop\_times} and replace each stop identifier with a geospatial H3 index derived from the stop coordinates. The resulting H3 sequence is the route-pattern key. Direction is encoded by order: the reverse direction produces a different sequence. Loops, short turns, deviations, and express patterns likewise produce distinct Route Patterns when their stop order differs.

This definition deliberately avoids using \texttt{shape\_id} or \texttt{stop\_id} as the Route Pattern identity. These identifiers are administrative fields, and they may change across GTFS releases even when the passenger-facing service is unchanged.

\paragraph{H3 Route Pattern Matching}

A critical challenge in multi-GTFS analysis is that stop IDs and shape IDs are administrative identifiers rather than stable passenger-facing Route Pattern identities. The same physical stop may receive different \texttt{stop\_id} values across feeds, and the same operational Route Pattern may receive multiple \texttt{shape\_id} values, even within one feed.

Barrie Transit Route~2A (Dunlop), obtained from MobilityDatabase \citep{mobilitydb2026} on February~20, 2026, provides a concrete example. The route operates as a single passenger-facing pattern serving 29 stops from Park Place to Downtown Barrie Terminal in direction~0, but its trips reference two \texttt{shape\_id} values. For reproducibility, the relevant GTFS identifiers are:

\begin{center}
\small
\begin{tabular}{ll}
\texttt{route\_id} & \texttt{c8bb5d6b-0a67-426c-8742-36d10b8c15b8} \\
\texttt{route\_short\_name} & \texttt{2A} \\
\texttt{route\_long\_name} & \texttt{DUNLOP} \\
\texttt{shape\_id} (58 trips) & \texttt{58ea5474-ef60-4dbc-bd74-5c122b012d32} \\
\texttt{shape\_id} (13 trips) & \texttt{c0134c50-192b-4a6c-8d93-d0534d3b0fab} \\
\end{tabular}
\end{center}

Both shapes contain 64 coordinate points.

\begin{figure}[H]
  \centering
  \includegraphics[width=0.38\textwidth]{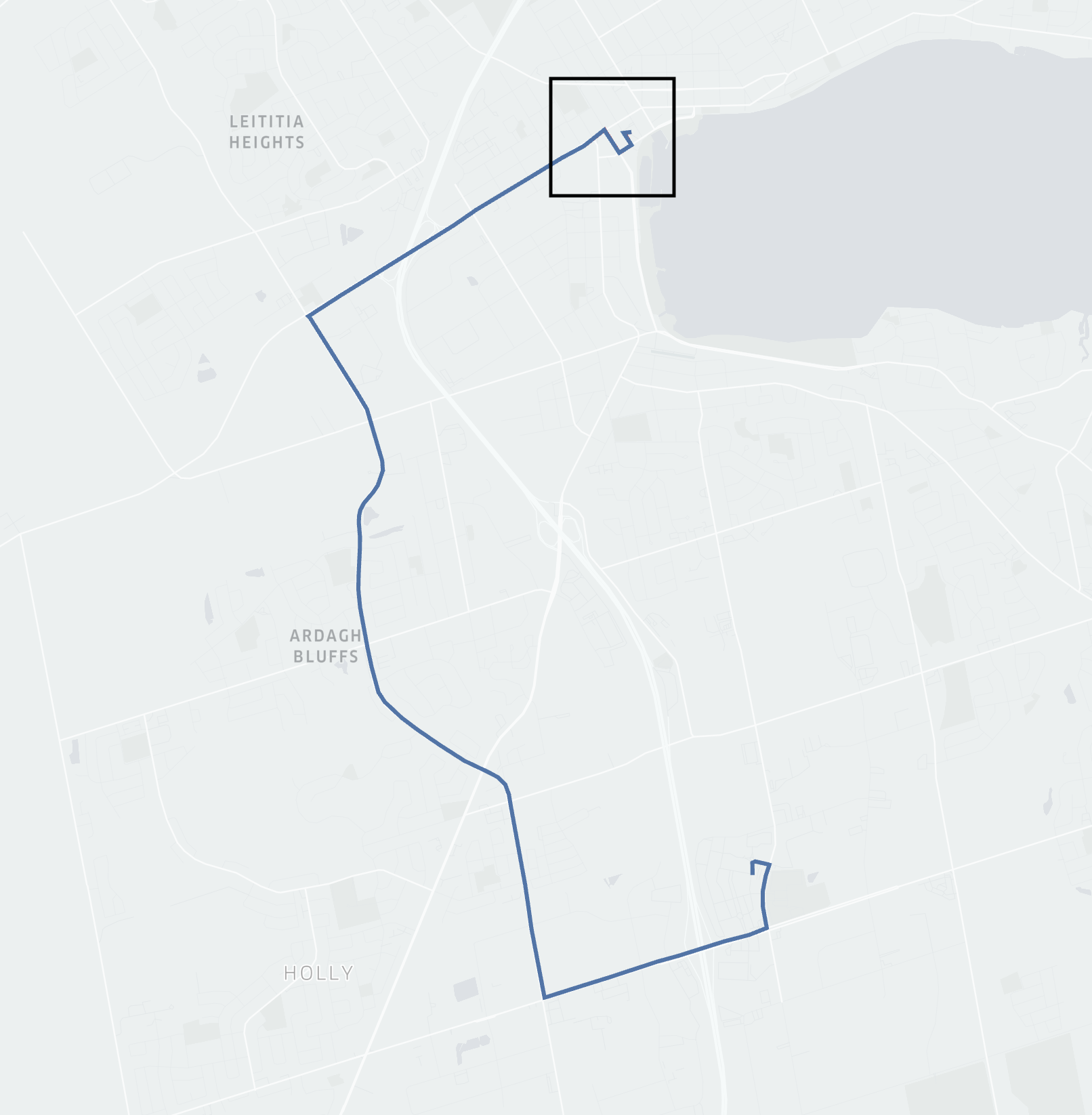}
  \hfill
  \includegraphics[width=0.58\textwidth]{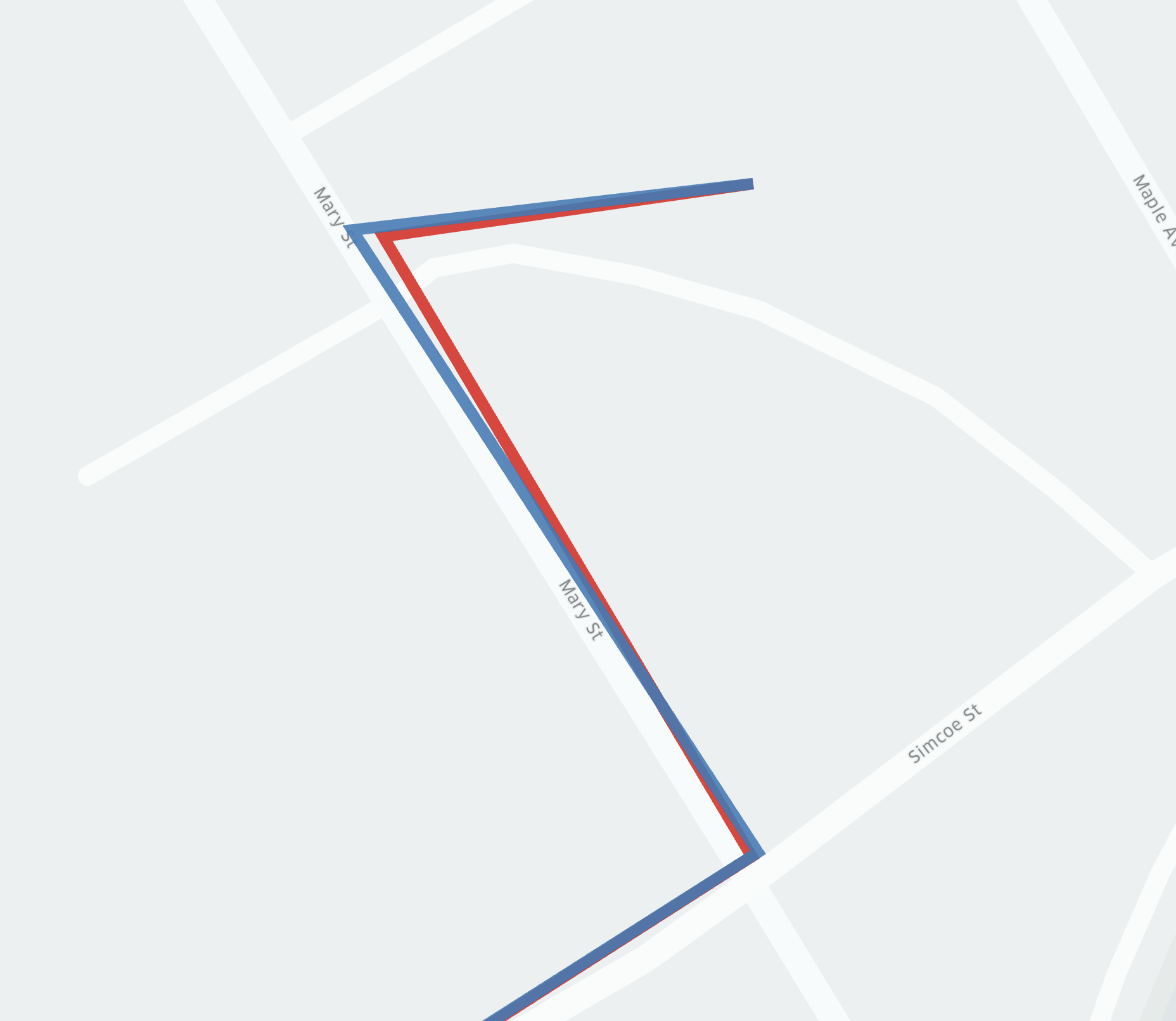}
  \caption{Two shape geometries assigned to the same Route~2A (Dunlop) pattern in Barrie Transit GTFS data. Left: full route scale---the polylines overlap completely. Right: close-up of the terminal approach near Downtown Barrie Terminal, showing the minor divergence (approximately 7.6~m) between the two \texttt{shape\_id} variants.}
  \label{fig:two_shapes}
\end{figure}

The two shapes are identical for the first 59 coordinate points and diverge only in the final terminal approach. Their total polyline length differs by 7.64~m over a route of nearly 11~km (0.07\%), while both serve the same 29 stops in the same order and terminate at the same location. Figure~\ref{fig:two_shapes} shows that the geometries overlap at route scale and differ only in the close-up terminal approach.

Using \texttt{shape\_id} as the Route Pattern identifier would split this single operational pattern into two. Dates with different mixes of the 58-trip and 13-trip shape assignments could then be classified as different DayTypes even though the passenger-visible service is unchanged. The H3 stop-sequence key avoids this artifact by grouping trips that share the same ordered geospatial stop sequence.

\paragraph{Resolving Identifier Instability}

We encode each stop location with the H3 hierarchical geospatial indexing system \citep{uber2018}, which has also been used in transit analysis for spatial clustering of GTFS data \citep{gramacki2021}. We use H3 resolution 15, which gives sub-meter-scale cells; coarser resolutions can be substituted when stop density or coordinate quality makes this too strict. A Route Pattern is then represented by the ordered sequence of H3 cells visited by its stops:

\begin{equation}
\text{RoutePatternKey}
=
\left(
\text{h3\_15}_{\text{stop}_1},
\text{h3\_15}_{\text{stop}_2},
\ldots,
\text{h3\_15}_{\text{stop}_n}
\right)
\end{equation}

where $\|$ denotes concatenation and $\text{h3\_15}_{\text{stop}_i}$ is the H3 resolution-15 cell containing stop $i$.

This sequence encodes direction because reverse directions have different ordered stop sequences. It also avoids dependence on \texttt{shape\_id}, \texttt{stop\_id}, or \texttt{direction\_id}, which may vary across feeds. The approach treats trips with the same stop sequence as the same Route Pattern even if their intermediate path geometry differs; spatially tolerant geometry matching is therefore left as a future extension for applications where inter-stop path differences are operationally important.

Returning to the Route~2A example from Barrie Transit: with H3-based matching, both shape variants produce the identical H3 sequence because all 29 stops are at the same physical locations. The approach correctly recognizes this as a single Route Pattern, eliminating the spurious distinction caused by different \texttt{shape\_id} values and ensuring accurate DayType computation.

\subsubsection{DayType Extraction from GTFS}
\label{sec:daytype_extraction_algorithm}

Having established Route Pattern extraction, we now extract complete DayTypes by aggregating all Route Pattern trips active on each date. Given a GTFS feed over date range $[d_{start},d_{end}]$, the algorithm returns the set of unique DayTypes $\mathcal{D}=\{\mathcal{D}_1,\ldots,\mathcal{D}_m\}$ and a mapping $\phi:[d_{start},d_{end}]\rightarrow\mathcal{D}$ assigning each date to one DayType.

The required GTFS inputs are \texttt{calendar}, \texttt{calendar\_dates}, \texttt{trips}, and \texttt{stop\_times}; \texttt{shapes} is optional for the present stop-sequence definition. The algorithm has three phases. First, weekly \texttt{calendar} records are expanded to date-level active service sets, and \texttt{calendar\_dates} exceptions are applied as service additions or removals. Second, for each active trip on each date, the ordered stop sequence is converted to an H3 Route Pattern key $K_t$, and the trip specification is formed as

\begin{equation*}
  \tau_t = (K_t, \text{first\_departure\_time}, \text{last\_arrival\_time}).
\end{equation*}

Third, all trip specifications active on date $d$ are collected into $\text{TripSet}_d$, and dates with identical trip-specification sets are assigned to the same DayType.

\paragraph{Computational Complexity}

Let $D$ be the number of dates, $S$ the number of service IDs, and $T$ the number of scheduled trips. Calendar expansion is $O(D \times S)$, trip collection is linear in the active trip records, and DayType grouping is a hash-based comparison of date-level trip sets. The overall procedure is $O(D \times T)$ in the conservative implementation used here and is linear in the size of the expanded date--trip representation.

The extraction and comparison procedures are released as the open-source Python package \texttt{gtfs\allowbreak-daytype} \citep{gtfsdaytype2026}. The package provides a reference implementation of H3 Route Pattern key construction, GTFS calendar expansion, DayType extraction, date-to-DayType calendar export, and exact and time-tolerant DayType comparison.

\subsubsection{Identification and Invariance Properties}

The extraction procedure is not only a data-processing pipeline; it defines the conditions under which DayTypes are identifiable from GTFS. Three properties are useful for methodological interpretation.

\medskip
\noindent\textbf{Invariance to GTFS identifier relabeling.}
Suppose two GTFS feeds describe the same passenger-facing timetable but differ only by one-to-one relabeling of \texttt{stop\_id}, \texttt{shape\_id}, \texttt{trip\_id}, or \texttt{service\_id}. If stop coordinates, stop order, service calendars, and stop times are unchanged, the H3 route-pattern keys and trip specifications are unchanged. Therefore the extracted DayType partition is unchanged.

\noindent\emph{Implication.} The method identifies service classes from geographic stop sequences and timetable content rather than from administrative identifiers. This property is essential for longitudinal GTFS analysis, where identifiers may change between feed releases without a corresponding change in passenger-facing service.

\medskip
\noindent\textbf{Exact DayType identifiability.}
Given a GTFS feed with valid calendar, exception, trip, and stop-time tables, the exact DayType for each date is uniquely determined by the set of active trip specifications $(p, dep, arr)$. Two dates are assigned to the same exact DayType if and only if their active trip-specification sets are equal.

\noindent\emph{Implication.} The exact extractor is identifiable in the deterministic sense: for a fixed feed and H3 resolution, there is a unique date-to-DayType mapping. This mapping is intentionally sensitive to any Route Pattern, trip-presence, departure-time, or arrival-time difference.

\medskip
\noindent\textbf{Bounded timing-shift stability.}
Let two DayTypes $A$ and $B$ contain the same number of trips, and suppose every trip in $A$ has a unique corresponding trip in $B$ with the same Route Pattern key and departure and arrival times differing by at most $\epsilon$ minutes. Then $I_{\epsilon}(A,B) = |A| = |B|$ and $d_{\epsilon}(A,B) = 0$.

\noindent\emph{Implication.} If two schedules have a complete one-to-one correspondence under the chosen tolerance, small timestamp shifts do not create dissimilarity under the tolerant metric, while the exact metric $d_0$ remains available as an audit signal. This gives the comparison framework a direct response to timetable noise without hiding planner-relevant timestamp differences.

\subsubsection{DayTypes as Route-Pattern-Time Schedule Sets}

The set in Equation~(3) is the Route-Pattern-time schedule set of a date. A DayType is exactly the equivalence class of dates that share this set. Therefore the schedule set is not a second object derived from the DayType; it is the timetable content that defines the DayType. Written as rows of a table, it has one row per scheduled trip and columns for Route Pattern, departure time, and arrival time. This representation is compact because many calendar dates map to the same DayType, but it remains detailed enough to be expanded into a date-specific timetable when required by downstream models.

This matters for the methodological literature reviewed below. Timetable synchronization, vehicle scheduling, deficit-function analysis, demand assignment, and activity-schedule models typically require a supply-side timetable as input, but public GTFS feeds do not state recurrent service classes directly. The proposed extractor identifies these recurring DayType schedule sets from GTFS before such models are solved. Date-specific deviations can then be represented as additions, removals, or timing shifts relative to the DayType schedule rather than as unrelated daily timetables. The exact metric $d_0$ identifies any deviation from the DayType schedule, while $d_{\epsilon}$ tests whether deviations remain after bounded same-Route-Pattern timing tolerance.

\section{Related Work}

This work sits at the boundary between transit data representation and the methodological literature on public transport planning, scheduling, and demand modeling. The key distinction is that we do not optimize a timetable, assign vehicles, or estimate activity schedules directly. Instead, we address the preceding representation problem: how to extract stable recurring service classes from GTFS so that such models can operate on a consistent service structure.

\textbf{Transit scheduling and timetable optimization.}
Recent research on transit scheduling and timetable optimization treats these problems as mathematical programs with explicit behavioral, operational, or capacity constraints. Liu and Ceder integrate public transport timetable synchronization, vehicle scheduling, and passenger demand assignment in a bi-objective bi-level model, using deficit-function structure to support scheduler decision-making \citep{liu-ceder-2018}. Their related retrospective shows how the deficit function has served as a bridge between public transport scheduling theory and practice for vehicle schedules, timetables, route networks, bus rapid transit, and crew duties \citep{liu-ceder-2017}. Zhang et~al. formulate cyclic train timetabling through an extended time-space network and solve the resulting reformulation using Lagrangian relaxation and ADMM methods \citep{zhang2019}. Cervantes-Sanmiguel et~al. study a bi-objective transit network design problem that trades off user travel time and monetary cost under operating-cost constraints \citep{cervantes2023}. These papers develop optimization models over timetable, network, or vehicle-allocation decision variables. Our paper supplies a complementary input layer: recurring DayType-specific supply classes extracted from GTFS and represented with stable Route Pattern keys.

\textbf{Demand and activity schedules.}
A second adjacent stream reconstructs demand-side temporal structure. Ballis and Dimitriou synthesize individual activity schedules from multi-period origin-destination matrices using graph-theoretical and combinatorial optimization concepts \citep{ballis2020}. Vo et~al. fuse household travel survey and smart-card data to generate spatiotemporally diverse activity schedules for transit users \citep{vo2026}. These methods recover or synthesize demand-side activity patterns. In contrast, our DayType extraction recovers the recurring supply-side service classes actually published by agencies. The two layers are naturally coupled: activity-schedule models describe when and where passengers need to travel, while DayType extraction identifies the recurring service supply against which those demand patterns can be assigned, compared, or optimized.

\textbf{GTFS, DayType representation, and transit analytics.}
GTFS is the de facto standard for transit schedule distribution \citep{mchugh2013}, but it does not encode DayTypes explicitly. Instead, service classes are implicit in \texttt{calendar}, \texttt{calendar\_dates}, \texttt{trips}, and \texttt{stop\_times}. The framework in this paper works within the existing GTFS schema: rather than requiring agencies to publish additional DayType tables, it reconstructs recurring service classes from data that agencies already provide. This avoids the adoption and consistency burden of maintaining additional denormalized files.

Beyond traditional scheduling, day type has emerged as a critical feature in modern transit analytics and machine learning applications. Recent work on real-time bus arrival prediction \citep{rashvand2023} demonstrates how day type classification (distinguishing weekdays from weekends) significantly improves the accuracy of neural network models for transit systems, with the day type variable directly incorporated into feature engineering for arrival time estimation. This practical application underscores the relevance of day type as a fundamental transit concept that bridges classical scheduling theory and modern data-driven analysis methods.

The contribution of the present paper is therefore deliberately different from, but connected to, the optimization and activity-schedule papers above. Scheduling and assignment models of this kind typically assume that the relevant timetable, service day, or recurrent service class is already available. In public GTFS data, that object is implicit, fragmented across files, and unstable across feed versions. We formalize and extract it, then define exact, time-tolerant, and structural-comparability metrics for comparing the resulting DayTypes. This closes a representation gap between published schedule data and downstream methodological models: the extracted DayType can serve as the recurrent supply object that is later expanded into time-space networks, deficit-function formulations, vehicle schedules, assignment models, or activity-schedule comparisons.

\section{Analysis Methods and Comparison Metrics}
\label{sec:analysis-methods}

Having established how to extract DayTypes from GTFS data (Section~\ref{sec:daytype_extraction_algorithm}), we now define metrics for comparing them. These metrics support schedule consolidation, change detection, service classification, anomaly detection, and cross-agency comparison.

\subsection{DayType Comparison Problem}

Given two DayTypes $A$ and $B$ from one feed, multiple feed versions, or different agencies, the comparison problem is to quantify whether they represent the same service, a subset relationship, a timing-shifted version of a comparable schedule, or a structural schedule difference.

\subsection{Comparison Metrics}

\subsubsection{Exact Set-Based Metrics}

Recall from Section~\ref{sec:definitions} that a DayType is a set of trip specifications:

\begin{equation}
\text{DayType}(d) = \{(p, dep_t, arr_t): \text{for all trips }t\text{ operating on }d\}.
\end{equation}

Given two DayTypes represented as sets $A = \text{DayType}(d_1)$ and $B = \text{DayType}(d_2)$, we define the symmetric difference distance.

\paragraph{Symmetric Difference Distance (Jaccard Distance)}
The exact Jaccard distance treats each trip specification as an atomic unit:

\begin{equation}
d_{\text{sym}}(A, B) = \frac{|A \triangle B|}{|A \cup B|} = \frac{|A \setminus B| + |B \setminus A|}{|A \cup B|}
\end{equation}

where $A \triangle B$ is the symmetric difference and $|A \cup B|$ is the total number of unique trip specifications.

The value lies in $[0,1]$, where 0 means identical DayTypes and 1 means no shared trip specifications. Because the metric is normalized by the union size, it is comparable across agencies with different service volumes.

\paragraph{Subset and Containment Relationships}
For cases where one DayType is a reduced version of another (e.g., weekend service as a subset of weekday service), we can define asymmetric measures:

\begin{equation}
\text{Containment}(A, B) = \frac{|A \cap B|}{|B|} \quad \text{(what fraction of B is contained in A?)}
\end{equation}

High containment indicates a subset relationship and suggests operationally related DayTypes.

\subsubsection{Time-Tolerant Timetable Matching}

The exact metrics above intentionally treat each trip specification as an atomic object:
\[
(h3\text{-}seq, dep, arr).
\]
This strict definition is useful for schedule auditing: if a trip that usually departs at 08:00 is encoded at 08:01 on another date, the exact comparison should flag that difference. Such shifts may reflect intentional timetable adjustment, a data-production artifact, or an unintended timetable inconsistency. Exact matching alone, however, cannot tell whether dissimilarity is caused by small timing shifts or by a substantive service redesign. We therefore add a time-tolerant comparison layer that keeps $d_0$ as a strict audit signal and asks how much of two DayTypes can be put into one-to-one correspondence after allowing bounded departure and arrival shifts.

Let a trip specification be denoted by $x = (p_x, dep_x, arr_x)$, where $p_x$ is the H3-based route-pattern key. For a tolerance parameter $\epsilon \geq 0$ measured in minutes, two trips are $\epsilon$-equivalent when they operate on the same Route Pattern and both their departure and arrival times differ by no more than $\epsilon$:

\begin{equation}
x \sim_{\epsilon} y
\iff
p_x = p_y
\land |dep_x - dep_y| \leq \epsilon
\land |arr_x - arr_y| \leq \epsilon .
\end{equation}

A useful special case is complete time-tolerant equivalence. Two DayTypes are equivalent at tolerance $\epsilon$ if every trip in $A$ can be paired with a unique $\epsilon$-equivalent trip in $B$, and every trip in $B$ is used in exactly one such pair. In that case the two DayTypes have the same Route Pattern trip structure at the chosen timing tolerance, even if the exact timestamps differ.

More generally, two DayTypes may be only partly matchable. Define $I_{\epsilon}(A,B)$ as the maximum count of unique $\epsilon$-equivalent trip pairs, with each trip used at most once. The time-tolerant Jaccard distance is:

\begin{equation}
d_{\epsilon}(A,B)
= 1 - \frac{I_{\epsilon}(A,B)}{|A| + |B| - I_{\epsilon}(A,B)} .
\end{equation}

The corresponding time-tolerant containment is:

\begin{equation}
C_{\epsilon}(A,B) = \frac{I_{\epsilon}(A,B)}{|B|}.
\end{equation}

When $\epsilon = 0$, these metrics reduce to the exact Jaccard distance and exact containment used above, assuming duplicate trip specifications have first been collapsed into set elements. A positive value of $\epsilon$ does not allow trips on different Route Patterns to match merely because their times are close. It only relaxes the timestamp component of the trip definition. The tolerance parameter therefore has a direct operational interpretation: it is the maximum departure and arrival shift allowed when asking whether two trips are corresponding trips.

The comparison between $d_0$ and $d_{\epsilon}$ is itself diagnostically useful. We define the strict--tolerant gap:

\begin{equation}
\Delta_{\epsilon}(A,B) = d_0(A,B) - d_{\epsilon}(A,B).
\end{equation}

A large $\Delta_{\epsilon}$ means that many exact differences disappear once bounded timing shifts are allowed. This does not make those differences irrelevant: it shows that exact dissimilarity is sensitive to small timestamp changes. Whether the pair should be interpreted as a shifted version of the same schedule depends on the matched share and trip-count imbalance defined below. Conversely, if both $d_0$ and $d_{\epsilon}$ remain high, the two DayTypes differ structurally even after accounting for small timing shifts.

\begin{table}[H]
\centering
\small
\renewcommand{\arraystretch}{1.12}
\begin{tabular}{@{}ll>{\raggedright\arraybackslash}p{0.50\textwidth}@{}}
\toprule
\textbf{Exact $d_0$} & \textbf{Tolerant $d_{\epsilon}$} & \textbf{Interpretation} \\
\midrule
Low  & Low  & Near-identical schedules \\
High & Low  & Timing-sensitive difference; check matched share and trip-count imbalance \\
Low  & High & Not possible for $\epsilon \geq 0$ under this definition \\
High & High & Substantive service difference or Route Pattern redesign \\
\bottomrule
\end{tabular}
\caption{Interpretation of the strict exact metric $d_0$ and time-tolerant metric $d_{\epsilon}$. The exact metric remains useful as an audit signal, while the tolerant metric tests whether exact differences persist after bounded timing tolerance.}
\label{tab:strict-tolerant-interpretation}
\end{table}

\paragraph{Interpretation of the tolerance parameter}
The tolerance parameter $\epsilon$ should be read as a sensitivity setting rather than as a replacement for exact comparison. At $\epsilon=0$, the metric is the strict schedule audit metric: a one-minute timestamp change is still visible. At $\epsilon=1$, $\epsilon=3$, or $\epsilon=5$, the comparison asks whether trips on the same Route Pattern can still be paired after allowing that many minutes of departure and arrival shift. The tolerance should remain small relative to the typical headway of the service being analyzed; otherwise distinct trips on the same Route Pattern may become ambiguous. In our empirical analysis, we therefore report one-, three-, and five-minute tolerances and interpret them together with matched share and trip-count imbalance.

\subsubsection{Structural Comparability and Timing Decomposition}
\label{sec:structural-comparability}

The tolerant metric $d_{\epsilon}$ is useful for detecting whether exact differences persist after bounded same-pattern timing shifts. It should not, however, be interpreted as evidence that two whole DayTypes are shifted versions of the same schedule when they have very different numbers of trips or operate different Route Patterns. In such cases, a few tolerance-based matches may coexist with large trip-presence differences. We therefore report timing tolerance together with structural-comparability diagnostics.

Let $A$ and $B$ be two DayTypes, and let $I_{\epsilon}(A,B)$ be the maximum time-tolerant match count defined above. The unmatched-trip count under tolerance $\epsilon$ is
\begin{equation}
U_{\epsilon}(A,B) = |A| + |B| - 2I_{\epsilon}(A,B),
\end{equation}
and the matched-trip share is
\begin{equation}
M_{\epsilon}(A,B) =
\frac{2I_{\epsilon}(A,B)}{|A|+|B|}.
\end{equation}
We also define the trip-count imbalance
\begin{equation}
Q(A,B) = \frac{||A|-|B||}{\max(|A|,|B|)} .
\end{equation}

These quantities separate two questions that the exact metric alone conflates. $d_{\epsilon}$ asks whether exact disjointness survives after bounded timing tolerance. $M_{\epsilon}$ and $U_{\epsilon}$ show how much of the two schedules is actually matched under that tolerance, while $Q$ identifies large service-level differences. A pair with low $d_{\epsilon}$, high $M_{\epsilon}$, and low $Q$ is plausibly a timing variation of a structurally comparable schedule. A pair with high $Q$ or low $M_{\epsilon}$ should instead be interpreted as a structural service difference, even if a small number of trips happen to match within tolerance.

\paragraph{Timing shifts for comparable pairs}
For structurally comparable pairs, the matched trips can be summarized by their timing shifts:
\begin{equation}
\bar{\delta}_{\epsilon}(A,B)
=
\frac{1}{I_{\epsilon}(A,B)}
\sum_{(x,y)\in \mathcal{M}_{\epsilon}^*(A,B)}
\frac{|dep_x-dep_y|+|arr_x-arr_y|}{2},
\end{equation}
where $\mathcal{M}_{\epsilon}^*(A,B)$ is a maximum-cardinality tolerant pairing. This statistic is reported only for the matched portion of the schedules and should be interpreted together with $M_{\epsilon}$ and $U_{\epsilon}$. It measures timing shifts among paired same-pattern trips; it does not imply that unmatched trips are shifted versions of one another.

\paragraph{Relation to the discrete metrics}
The comparison hierarchy is therefore: exact audit first, tolerant matching second, and structural-comparability diagnostics third. The exact metric $d_0$ asks whether the trip-specification sets are identical. The tolerant metric $d_{\epsilon}$ asks whether exact differences persist after bounded same-pattern timing shifts. The matched share, unmatched count, and trip-count imbalance then determine whether it is appropriate to describe the remaining difference as timing variation of a comparable schedule or as structural service change.

\paragraph{Pattern-level limits and future extensions}
The current definition treats H3-based pattern keys as the Route Pattern equivalence relation. Genuine geometric or stop-sequence changes that produce different keys for functionally equivalent service are therefore interpreted as structural differences. A spatially tolerant Route Pattern matching---relaxing exact H3-sequence equality, sampling intermediate path geometry, or applying trajectory-based distances such as SSPD~\citep{besse2016}---would extend the framework to distinguish small geometry changes from true Route Pattern redesign, and is reserved for future work.

\section{Results}

We applied the method to public GTFS feeds from Japan and Canada to test extraction across different network scales, calendar conventions, and scheduling practices.

\subsection{Case Study 1: Otone Kotsu (Japan)}

Otone Kotsu is a regional operator in Kanto, Japan. Its feed covers March~17, 2025 through May~5, 2026 (415 days), with 11 routes, 109 stops, and 36 H3 Route Patterns. The extractor identifies four DayTypes and assigns every date: DT3 is the weekday pattern (239 days, 128 trips), DT0 is the weekend-and-holiday pattern (122 days, 78 trips), DT1 is the January~1--3 \emph{sanganichi} New Year pattern (3 days, 48 trips), and DT2 is a single Saturday exception (June~7, 2025; 93 trips).

The metrics distinguish these exception types. DT2 is close to the weekend/holiday schedule, with $d(\text{DT0}, \text{DT2})=0.161$ and $C(\text{DT0}, \text{DT2})=0.957$, indicating a Saturday-like schedule with additional trips. DT1 is far from every other DayType ($d>0.94$), indicating a distinct reduced-service regime. The holiday assignments include Japanese national holidays defined by the Cabinet Office of Japan~\citep{cao-jp}; the extraction itself does not use this external list.

\begin{figure}[H]
  \centering
  \includegraphics[width=\textwidth]{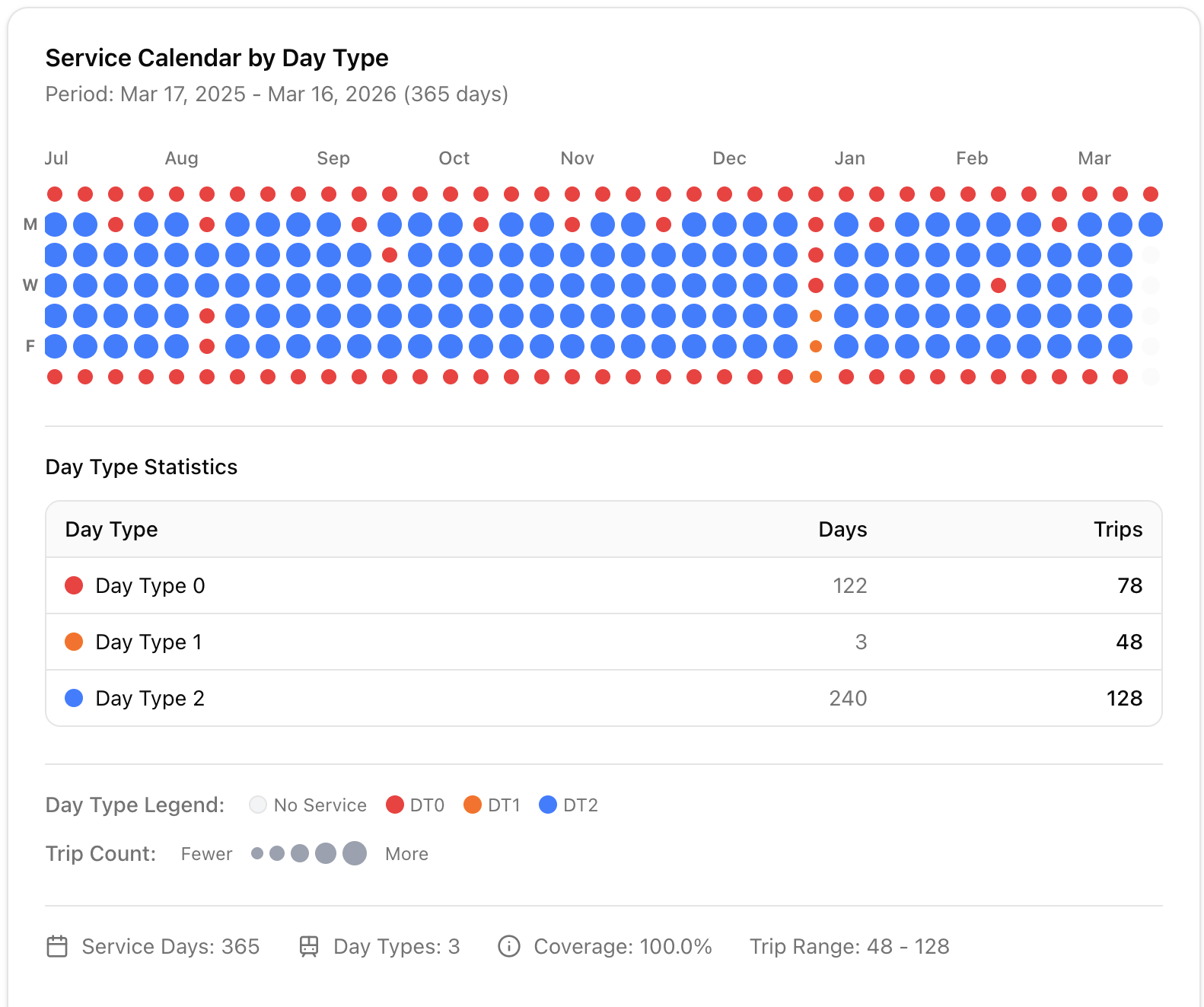}
  \caption{Service Calendar by DayType for Otone Kotsu (fragment showing July 2025--March 2026). Each column is a week, each row is a day of the week, and colours distinguish the four DayTypes. The weekday, weekend/holiday, New Year, and single-day exception patterns are separated without external calendar labels.}
  \label{fig:otone}
\end{figure}

\subsection{Case Study 2: Barrie Transit (Canada)}

Barrie Transit provides a compact North American comparison case. The Mobility Database feed (mdb-3) covers January~12 through April~6, 2024 (84 days) \citep{barrie2024}; the network contains 20 routes, 656 stops, and 37 H3 Route Patterns. The extractor identifies a clean three-tier weekly cycle: Sunday (DT0, 11 days, 307 trips), weekday (DT1, 60 days, 658 trips), and Saturday (DT2, 13 days, 571 trips). Saturday service reaches 86.8\% of weekday trip volume, while Sunday service reaches 46.7\%.

\begin{figure}[H]
  \centering
  \includegraphics[width=\textwidth]{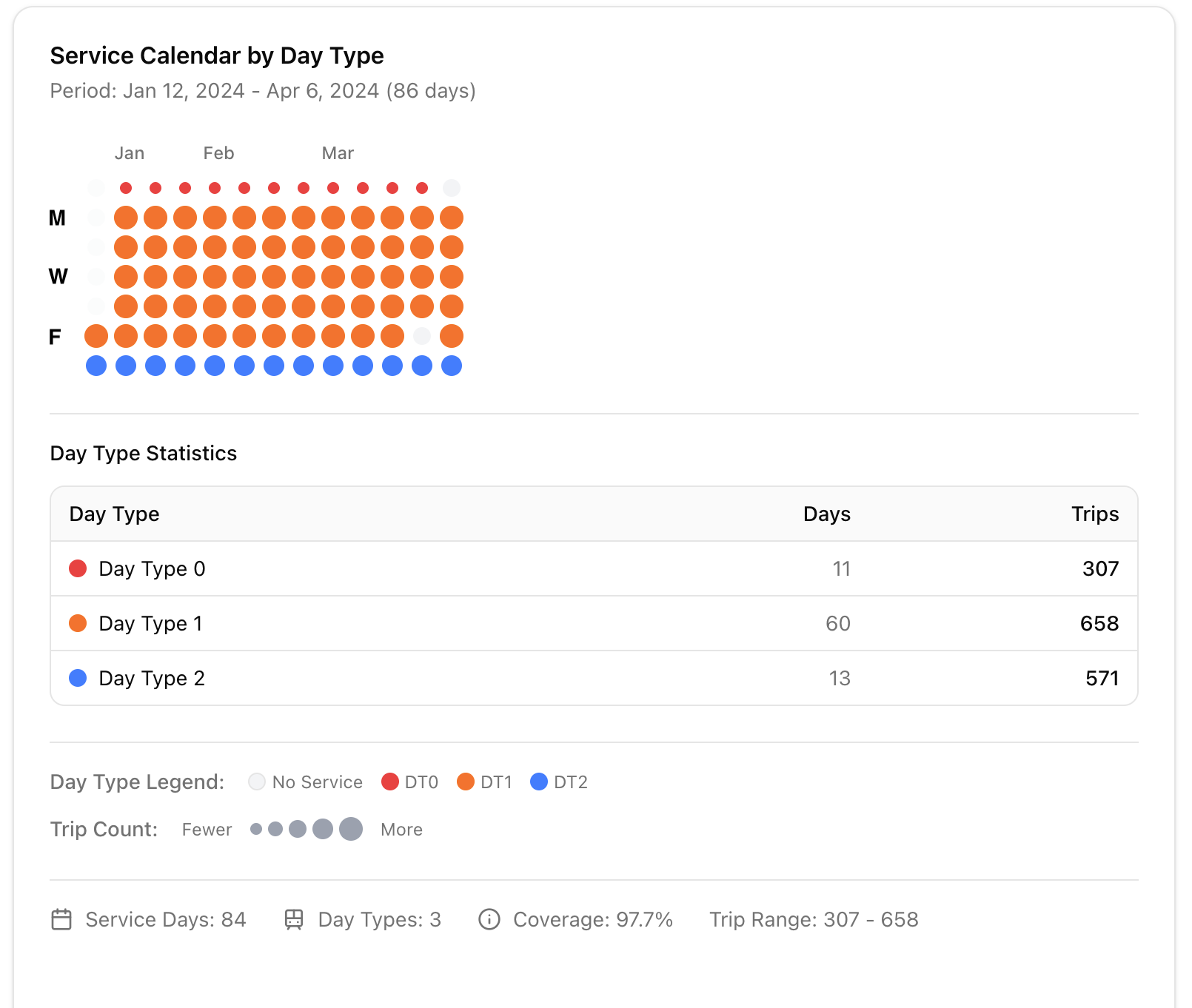}
  \caption{Service Calendar by DayType for Barrie Transit, January--April 2024. The three-color scheme corresponds to weekdays, Saturdays, and Sundays.}
  \label{fig:barrie}
\end{figure}

\subsection{Case Study 3: Halifax Transit (Canada)}

Halifax Transit provides a larger Canadian case. The Mobility Database feed (mdb-734) covers February~23 through May~17, 2026 (84 days) \citep{halifax2026}; the network contains 80 routes, 2{,}381 stops, and 181 H3 Route Patterns. The extractor identifies five DayTypes: weekday (DT0, 59 days, 4{,}200 trips), Saturday (DT1, 12 days, 2{,}765 trips), Sunday (DT2, 11 days, 2{,}217 trips), and two single-day reduced-service configurations (DT3 and DT4, 2{,}147--2{,}213 trips). Unlike Barrie, Halifax therefore combines the standard weekly cycle with isolated exception days.

\begin{figure}[H]
  \centering
  \includegraphics[width=\textwidth]{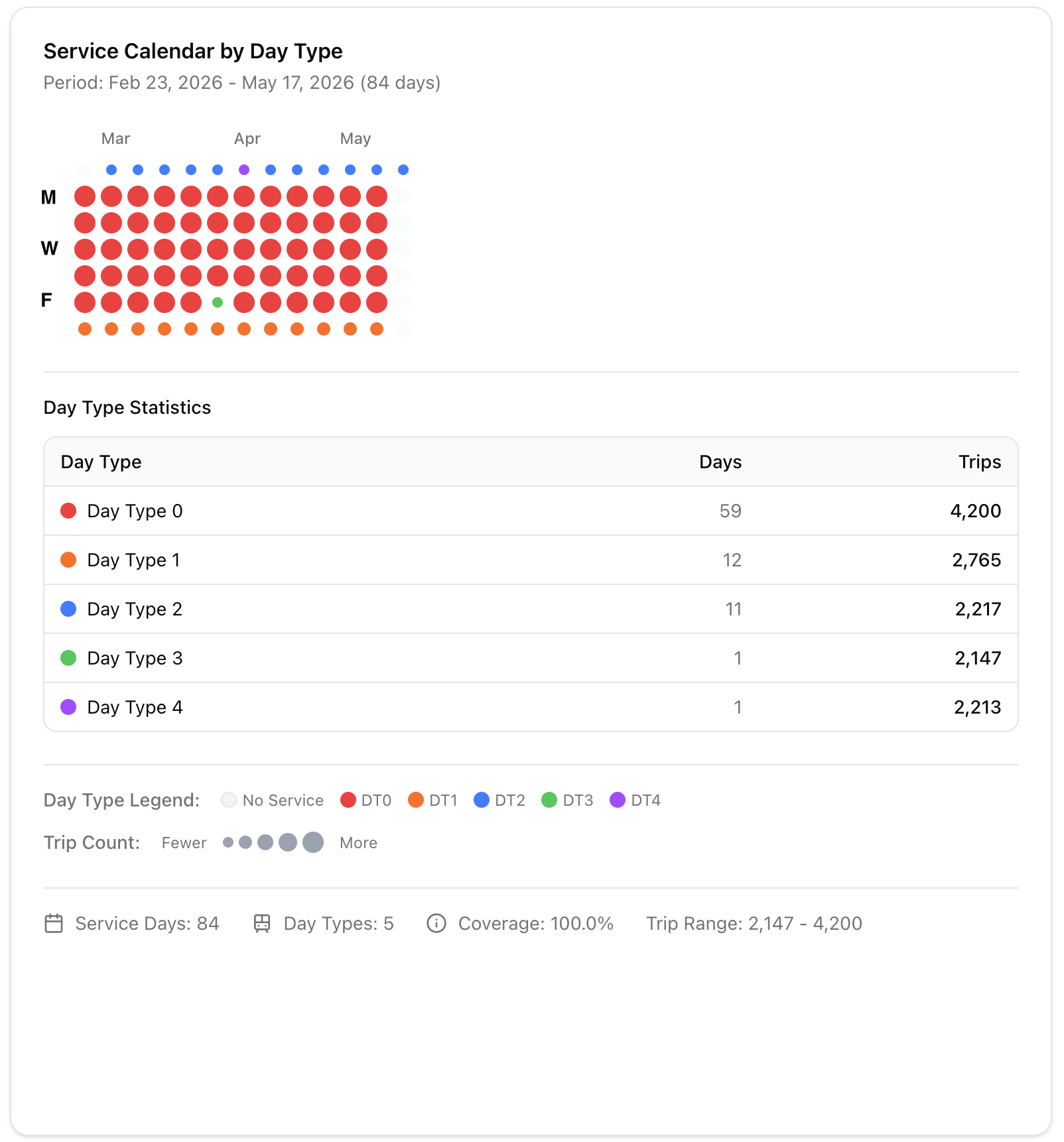}
  \caption{Service Calendar by DayType for Halifax Transit, February--May 2026. The five-color scheme shows weekday, Saturday, Sunday, and two single-day reduced-service configurations.}
  \label{fig:halifax}
\end{figure}

\subsection{Comparative Analysis}

The three case studies span two countries, three network scales, and different calendar-encoding strategies. Barrie uses a weekday--Saturday--Sunday structure; Otone Kotsu adds holiday and New Year exceptions; and Halifax adds two single-day reduced-service configurations. In all cases the extractor assigns 100\% of dates without manual intervention, supporting the claim that DayTypes can be reconstructed from standard GTFS feeds despite different calendar encodings.

\section{Cross-National DayType Distance Analysis}
\label{sec:cross-national}

To move beyond individual case studies, we applied the extraction algorithm and distance metrics to publicly available GTFS feeds from 40 transit agencies in Japan and 38 land-based transit agencies in Canada. Japanese feeds were collected from the Open Data Portal for Public Transportation (ODPT)~\citep{odpt2026}; Canadian feeds were obtained from the Canadian Public Transit Network Database published by Statistics Canada~\citep{statcan2025}, accessed on April 3, 2026. For each agency, we analyzed the most recent GTFS file and computed pairwise Jaccard distances and containment measures between all extracted nonempty DayTypes.

We exclude two ferry operators, BC Ferries and Aquabus, because water-based transit has distinct operational drivers: tides, weather-dependent sailings, seasonal routing, and vessel constraints. Applied to these feeds, the algorithm detects 66 DayTypes for BC Ferries and 1 degenerate DayType for Aquabus, confirming that ferry schedules deserve separate analysis.

\subsection{Summary Statistics}

For each agency, we computed the number of extracted DayTypes and pairwise exact Jaccard/containment summaries: the minimum Jaccard distance $d_{\min}$ (the closest pair of DayTypes), the mean pairwise distance $\bar{d}$, the maximum distance $d_{\max}$ (the most dissimilar pair), and the corresponding containment statistics. Pairwise summaries require at least two nonempty DayTypes; after this filter, the aggregate comparison includes 39 Japanese agencies and 33 Canadian agencies. Tables~\ref{tab:japan-summary} and~\ref{tab:canada-summary} report representative agency-level summaries, with full pairwise distance and containment matrices provided as supplementary material.

\begin{table}[H]
\centering
\scriptsize
\resizebox{\textwidth}{!}{%
\begin{tabular}{lrrrrrrrrrr}
\hline
\textbf{Agency} & \textbf{Rts} & \textbf{Stops} & \textbf{Trips} & \textbf{DTs} & $d_{\min}$ & $\bar{d}$ & $d_{\max}$ & $C_{\min}$ & $\bar{C}$ & $C_{\max}$ \\
\hline
Keifuku Bus              & 68  & 1{,}566 & 1{,}350  & 39 & .004 & .352 & .697 & .385 & .752 & 1.00 \\
Gunma Bus                & 44  & 1{,}210 & 1{,}468  & 33 & .002 & .246 & .708 & .332 & .850 & 1.00 \\
Kan-etsu Transportation  & 73  & 2{,}262 & 2{,}965  & 26 & .002 & .397 & .813 & .237 & .727 & 1.00 \\
Kyoto City Bus           & 141 & 1{,}673 & 18{,}553 & 16 & .001 & .357 & .632 & .491 & .747 & 1.00 \\
Akiha Bus Service        & 5   & 343     & 601      & 13 & .012 & .478 & .920 & .115 & .591 & 1.00 \\
Yokohama Municipal Bus   & 149 & 2{,}517 & 22{,}325 & 12 & .062 & .783 & .990 & .012 & .370 & 1.00 \\
Nippon Chuo Bus (Maebashi) & 36 & 590   & 983      & 9  & .011 & .336 & .594 & .521 & .769 & 1.00 \\
Nagai Transportation     & 24  & 401     & 491      & 7  & .013 & .459 & .724 & .371 & .670 & 1.00 \\
Akita City Bus           & 53  & 750     & 535      & 6  & .020 & .356 & .584 & .515 & .760 & 1.00 \\
Aomori City              & 288 & 834     & 4{,}836  & 6  & .009 & .617 & .996 & .008 & .397 & 1.00 \\
Hokkaido Takushoku (Hwy) & 7   & 107     & 48       & 6  & .023 & .058 & .104 & .896 & .971 & 1.00 \\
Hokkaido Takushoku (Reg) & 32  & 880     & 459      & 6  & .009 & .360 & .576 & .458 & .787 & 1.00 \\
Daishinto Radiant City   & 3   & 9       & 456      & 5  & .106 & .602 & .816 & .245 & .544 & 1.00 \\
Hitachi Auto (Chiyoda)   & 6   & 79      & 91       & 4  & .164 & .711 & 1.00 & .000 & .333 & 1.00 \\
Oshima Bus               & 23  & 101     & 51       & 4  & .020 & .106 & .157 & .843 & .947 & 1.00 \\
Otone Kotsu              & 11  & 109     & 347      & 4  & .161 & .831 & .982 & .031 & .216 & 1.00 \\
Shimoden Bus             & 22  & 641     & 583      & 4  & .025 & .549 & .761 & .342 & .579 & 1.00 \\
Gunma Chuo Bus           & 11  & 369     & 267      & 3  & .697 & .777 & .841 & .159 & .552 & 1.00 \\
Joshin Kanko Bus         & 2   & 101     & 165      & 3  & .098 & .135 & .197 & .803 & .933 & 1.00 \\
Nishitokyo City          & 7   & 146     & 669      & 3  & .536 & .625 & .697 & .414 & .550 & .658 \\
Tokyo Chuo City          & 2   & 73      & 147      & 3  & .069 & .690 & 1.00 & .000 & .322 & 1.00 \\
\hline
\multicolumn{11}{l}{\textit{18 additional agencies omitted for brevity; full matrices are provided as supplementary material.}} \\
\hline
\end{tabular}%
}
\caption{Representative DayType distance metrics for Japanese transit agencies, sorted by number of extracted DayTypes. $d$: Jaccard distance; $C$: containment.}
\label{tab:japan-summary}
\end{table}

\begin{table}[H]
\centering
\scriptsize
\resizebox{\textwidth}{!}{%
\begin{tabular}{lrrrrrrrrrr}
\hline
\textbf{Agency} & \textbf{Rts} & \textbf{Stops} & \textbf{Trips} & \textbf{DTs} & $d_{\min}$ & $\bar{d}$ & $d_{\max}$ & $C_{\min}$ & $\bar{C}$ & $C_{\max}$ \\
\hline
BC Tr.\ West Kootenay    & 34  & 637     & 1{,}038  & 17 & .005 & .371 & 1.00 & .000 & .677 & 1.00 \\
BC Tr.\ East Kootenay    & 13  & 249     & 1{,}148  & 13 & .011 & .628 & 1.00 & .000 & .452 & 1.00 \\
BC Tr.\ Revelstoke       & 3   & 89      & 300      & 10 & .103 & .425 & .765 & .235 & .763 & 1.00 \\
BC Tr.\ Fraser Valley    & 37  & 949     & 3{,}748  & 8  & .002 & .451 & .979 & .030 & .568 & 1.00 \\
BC Tr.\ North Okanagan   & 20  & 407     & 1{,}332  & 8  & .011 & .569 & .909 & .113 & .543 & 1.00 \\
BC Tr.\ Victoria         & 67  & 2{,}348 & 12{,}620 & 7  & .001 & .662 & .991 & .014 & .352 & 1.00 \\
BC Tr.\ Comox Valley     & 16  & 263     & 517      & 6  & .015 & .597 & 1.00 & .000 & .437 & 1.00 \\
BC Tr.\ Powell River     & 7   & 244     & 237      & 6  & .057 & .612 & 1.00 & .000 & .447 & 1.00 \\
BC Tr.\ Hazelton         & 20  & 293     & 552      & 5  & .044 & .535 & .770 & .232 & .673 & 1.00 \\
BC Tr.\ Nanaimo          & 22  & 843     & 2{,}249  & 5  & .007 & .688 & .991 & .013 & .332 & .997 \\
BC Tr.\ Prince George    & 18  & 563     & 1{,}023  & 5  & .012 & .539 & .866 & .156 & .597 & 1.00 \\
BC Tr.\ Quesnel          & 8   & 141     & 202      & 5  & .071 & .338 & .696 & .375 & .768 & 1.00 \\
BC Tr.\ S.\ Okanagan     & 16  & 374     & 400      & 5  & .009 & .507 & .975 & .028 & .588 & 1.00 \\
BC Tr.\ Cowichan Valley  & 15  & 602     & 503      & 4  & .112 & .840 & 1.00 & .000 & .183 & 1.00 \\
BC Tr.\ Creston          & 3   & 32      & 53       & 4  & .125 & .317 & .474 & .625 & .822 & 1.00 \\
BC Tr.\ Fort St.\ John   & 5   & 94      & 353      & 4  & .110 & .808 & 1.00 & .000 & .244 & .942 \\
BC Tr.\ Kamloops         & 26  & 591     & 1{,}823  & 4  & .006 & .824 & 1.00 & .000 & .186 & .998 \\
BC Tr.\ Kelowna          & 32  & 975     & 2{,}642  & 4  & .003 & .828 & .996 & .006 & .179 & 1.00 \\
BC Tr.\ Salt Spring Isl. & 7   & 27      & 233      & 4  & .320 & .798 & .922 & .127 & .296 & .809 \\
BC Tr.\ Squamish         & 5   & 122     & 283      & 4  & .031 & .833 & 1.00 & .000 & .175 & 1.00 \\
BC Tr.\ Sunshine Coast   & 5   & 297     & 342      & 4  & .044 & .739 & 1.00 & .000 & .291 & .992 \\
Brandon Transit          & 13  & 272     & 1{,}060  & 4  & .360 & .572 & .748 & .275 & .662 & 1.00 \\
Barrie Transit           & 13  & 529     & 1{,}511  & 3  & .758 & .860 & .924 & .104 & .254 & .405 \\
BC Tr.\ Campbell River   & 8   & 284     & 378      & 3  & .591 & .751 & .853 & .178 & .423 & .767 \\
BC Tr.\ Merritt          & 3   & 68      & 138      & 3  & .170 & .387 & .541 & .475 & .790 & 1.00 \\
BC Tr.\ Mt Waddington    & 7   & 72      & 48       & 3  & 1.00 & 1.00 & 1.00 & .000 & .000 & .000 \\
BC Tr.\ Port Alberni     & 4   & 145     & 228      & 3  & .951 & .984 & 1.00 & .000 & .032 & .102 \\
BC Tr.\ Prince Rupert    & 8   & 117     & 289      & 3  & .140 & .279 & .395 & .632 & .849 & 1.00 \\
BC Tr.\ Williams Lake    & 4   & 77      & 55       & 3  & .075 & .692 & 1.00 & .000 & .321 & 1.00 \\
Belleville Transit       & 10  & 221     & 1{,}196  & 3  & .341 & .615 & .784 & .286 & .543 & .893 \\
Burlington Transit       & 16  & 789     & 2{,}602  & 3  & .240 & .716 & .960 & .056 & .359 & .983 \\
BC Tr.\ Clearwater       & 4   & 19      & 14       & 2  & .286 & .286 & .286 & .833 & .833 & .833 \\
BC Tr.\ Smithers         & 2   & 20      & 32       & 2  & .968 & .968 & .968 & .050 & .067 & .083 \\
BC Tr.\ 100 Mile House   & 2   & 22      & 8        & 1  & ---  & ---  & ---  & ---  & ---  & ---  \\
BC Tr.\ Ashcroft         & 1   & 4       & 4        & 1  & ---  & ---  & ---  & ---  & ---  & ---  \\
BC Tr.\ Dawson Creek     & 2   & 74      & 20       & 1  & ---  & ---  & ---  & ---  & ---  & ---  \\
BC Tr.\ Pemberton Valley & 2   & 37      & 27       & 1  & ---  & ---  & ---  & ---  & ---  & ---  \\
BC Tr.\ Whistler         & 13  & 152     & 696      & 1  & ---  & ---  & ---  & ---  & ---  & ---  \\
\hline
\end{tabular}%
}
\caption{DayType distance metrics for Canadian transit agencies; ferry operators excluded. $d$: Jaccard distance; $C$: containment.}
\label{tab:canada-summary}
\end{table}

\subsection{Aggregate Findings}

Table~\ref{tab:aggregate} summarizes the key pairwise metrics across both countries. The extraction sample contains 78 GTFS feeds (40 Japanese and 38 Canadian, with ferry operators excluded from Canada). The aggregate pairwise summaries are computed over agencies with at least two nonempty DayTypes: 39 in Japan and 33 in Canada.

\begin{table}[H]
\centering
\small
\renewcommand{\arraystretch}{1.12}
\begin{tabular}{@{}>{\raggedright\arraybackslash}p{0.52\textwidth}cc@{}}
\toprule
\textbf{Metric} & \textbf{Japan} & \textbf{Canada} \\
 & \textbf{(39 agencies)} & \textbf{(33 agencies)} \\
\midrule
Median DayTypes per agency & 4 & 4 \\
Mean DayTypes per agency & 8.0 & 5.1 \\
Maximum DayTypes per agency & 39 & 17 \\
\addlinespace[2pt]
\multicolumn{3}{@{}l}{\textit{Jaccard distance, averaged across agencies}} \\
Closest pair, $\bar{d}_{\min}$ & 0.204 & 0.211 \\
Mean pairwise distance, $\bar{d}$ & 0.509 & 0.637 \\
Most distant pair, $\bar{d}_{\max}$ & 0.698 & 0.873 \\
\addlinespace[2pt]
\multicolumn{3}{@{}l}{\textit{Containment, averaged across agencies}} \\
Minimum containment, $\bar{C}_{\min}$ & 0.349 & 0.147 \\
Mean containment, $\bar{C}$ & 0.585 & 0.446 \\
Maximum containment, $\bar{C}_{\max}$ & 0.872 & 0.873 \\
\addlinespace[2pt]
\multicolumn{3}{@{}l}{\textit{Structural patterns}} \\
Agencies with full nesting, $C_{\max}=1.0$ & 27/39 (69\%) & 19/33 (58\%) \\
Agencies with exact disjointness, $d_{\max}=1.0$ & 6/39 (15\%) & 12/33 (36\%) \\
\bottomrule
\end{tabular}
\caption{Aggregate DayType distance statistics by country. Statistics computed over agencies with at least two nonempty DayTypes.}
\label{tab:aggregate}
\end{table}

The national maxima are driven by Keifuku Bus in Japan (39 exact DayTypes) and BC Transit West Kootenay in Canada (17 exact DayTypes). These should be read as exact schedule fragmentation rather than as necessarily distinct planner-facing service regimes: the closest DayType pairs have $d_{\min}=0.004$ for Keifuku and $d_{\min}=0.005$ for West Kootenay, indicating that some exact DayTypes differ by only a small number of trip specifications.

\paragraph{Aggregate patterns}
Both countries have a median of four DayTypes per agency, but Japan has more high-DayType outliers (maximum 39 vs.\ 17 in Canada), consistent with finer-grained holiday, seasonal, and exception calendars in some Japanese feeds. Canadian agencies have higher maximum pairwise distances ($\bar{d}_{\max}=0.87$ vs.\ $0.70$) and more complete exact disjointness (36\% vs.\ 15\%), whereas Japanese agencies show slightly more full nesting (69\% vs.\ 58\%). Closest-pair distances are similar ($\bar{d}_{\min}=0.204$ in Japan and $0.211$ in Canada), indicating that both samples include near-duplicate DayTypes that may reflect minor weekday variants or holiday-adjacent schedules.

\paragraph{Zero containment: complete disjointness between DayTypes}
A particularly informative finding is the occurrence of $C_{\min} = 0$---pairs of DayTypes that share \emph{no trips at all}. Before analysing these cases, we note that some GTFS feeds contain \emph{degenerate DayTypes} with zero trips: the calendar defines service windows on certain dates but assigns no trips to them (e.g., Yokohama Municipal Bus and Nagoya SRT in our dataset). We exclude such 0-trip DayTypes from all distance and containment calculations, as they represent a data encoding choice (explicitly marking ``no service'' days) rather than an operational pattern. All metrics reported in this paper reflect this exclusion.

After filtering, zero containment still occurs in 6 of 39 Japanese agencies (15\%) and 12 of 33 Canadian agencies (36\%). The exact metric alone would classify all of these cases as complete disjointness, but that label can arise from different operational mechanisms. The strict--tolerant framework separates two questions: whether any trip specifications are repeated exactly, and whether the same Route Pattern service remains after allowing small timing shifts.

\emph{(i) Timing shifts on shared Route Patterns.} Some exact-disjoint pairs retain much of the same Route Pattern structure but use different times or service volumes. BC Transit Kamloops illustrates this: its Saturday DayType (587 trips) shares zero exact trip specifications with either weekday DayType (845--846 trips), yielding $d_0=1.0$ and $C_0=0$, yet 33 of 35 weekday Route Pattern identifiers also appear on Saturday. The tolerant metric tests whether this exact disjointness survives after allowing small same-pattern timing shifts; the matched share then determines whether the result reflects broad timing variation or only limited overlap.

We therefore recomputed all exactly disjoint DayType pairs under one-, three-, and five-minute tolerances. Table~\ref{tab:exact-disjoint-tolerance} reports the share of exactly disjoint pairs that remain completely disjoint after tolerance is applied. In Canada, 43.4\% of exact-disjoint pairs remain disjoint after a one-minute tolerance, and 39.6\% remain disjoint after five minutes. Equivalently, more than half of the Canadian exact-disjoint pairs acquire at least one same-pattern match once a one-minute tolerance is allowed. In Japan, the corresponding remaining shares are higher (86.4\% at one minute and 59.1\% at five minutes), indicating that exact disjointness there is more often persistent under small timing tolerances. The empirical comparison therefore changes the interpretation of the original exact metric: exact disjointness is not discarded, but decomposed into pairs with some tolerant same-pattern overlap and pairs that remain completely unmatched.

\begin{table}[H]
\centering
\scriptsize
\resizebox{\textwidth}{!}{%
\begin{tabular}{lrrrr}
\hline
\textbf{Country} & \textbf{Exact-disjoint pairs} & \textbf{Remain disjoint at 1 min} & \textbf{3 min} & \textbf{5 min} \\
\hline
Japan  & 22 & 86.4\% & 63.6\% & 59.1\% \\
Canada & 53 & 43.4\% & 43.4\% & 39.6\% \\
\hline
\end{tabular}%
}
\caption{Persistence of exact disjointness under time-tolerant matching. The table conditions on DayType pairs with $d_0=1$ and reports the percentage that still have $d_{\epsilon}=1$ after allowing same-pattern trips to match within $\epsilon$ minutes of both departure and arrival time.}
\label{tab:exact-disjoint-tolerance}
\end{table}

The tolerance results do not by themselves imply that exact-disjoint pairs are shifted versions of the same schedule. A pair may acquire a few same-pattern matches under tolerance while still differing in most of its trips. We therefore summarize the same exact-disjoint pairs using the structural-comparability diagnostics introduced in Section~\ref{sec:structural-comparability}. Table~\ref{tab:structural-comparability-disjoint} reports the median number of five-minute tolerant matches, the median matched-trip share $M_5$, the median trip-count imbalance $Q$, and the median unmatched-trip count $U_5$.

\begin{table}[H]
\centering
\scriptsize
\resizebox{\textwidth}{!}{%
\begin{tabular}{lrrrrr}
\hline
\textbf{Country} & \textbf{Pairs} & \textbf{Median matches at 5 min} & \textbf{Median $M_5$} & \textbf{Median $Q$} & \textbf{Median $U_5$} \\
\hline
Japan  & 22 & 0 & 0.0\% & 85.6\% & 155 \\
Canada & 53 & 8 & 6.3\% & 70.9\% & 190 \\
\hline
\end{tabular}%
}
\caption{Structural comparability of exactly disjoint DayType pairs after five-minute time-tolerant matching. $M_5$ is the matched-trip share, $Q$ is trip-count imbalance, and $U_5$ is the unmatched-trip count after tolerant matching. Low $M_5$ and high $Q$ indicate structural service differences rather than timing variation of a comparable schedule.}
\label{tab:structural-comparability-disjoint}
\end{table}

This decomposition changes the interpretation of Table~\ref{tab:exact-disjoint-tolerance}. In Canada, many exact-disjoint pairs cease to be completely disjoint once tolerance is allowed, but the median matched share at five minutes is only 6.3\%, and the median trip-count imbalance is 70.9\%. Thus tolerance often reveals limited same-pattern overlap, not whole-schedule timing variation. In Japan, the median matched share is 0.0\% and the median trip-count imbalance is 85.6\%, indicating even stronger structural separation. Exact, tolerant, and structural-comparability statistics should therefore be reported together: $\Delta_{\epsilon}$ identifies whether any same-pattern timing overlap exists, while $M_{\epsilon}$, $U_{\epsilon}$, and $Q$ determine whether that overlap is large enough to interpret as a timing-shifted comparable schedule rather than structural service change.

\subsection{Illustrative Distance Matrices}

\begin{figure}[H]
  \centering
  \includegraphics[width=0.95\textwidth]{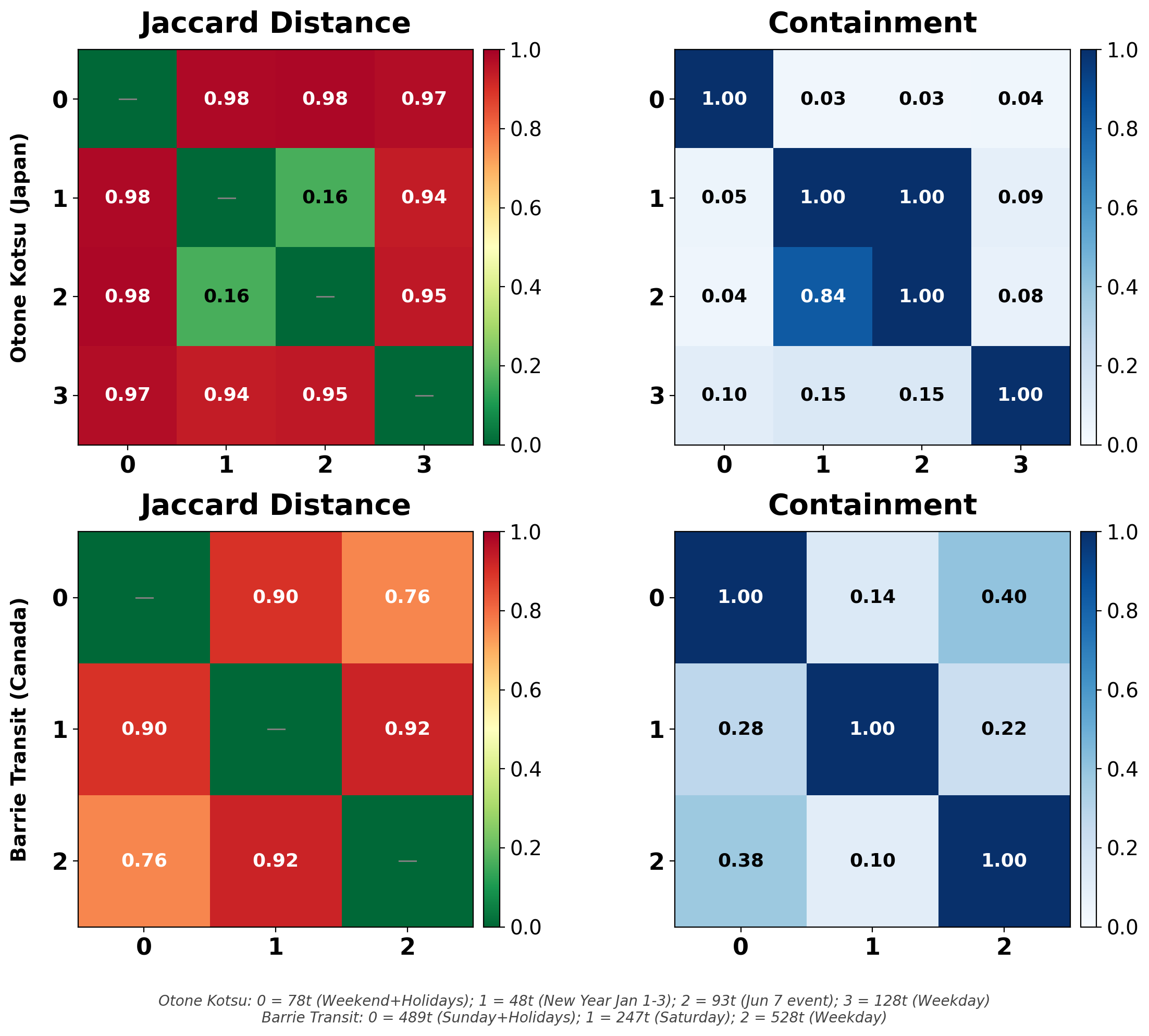}
  \caption{Jaccard distance (left) and containment (right) matrices for the two case-study agencies present in our cross-national dataset. Green/yellow cells indicate similar DayTypes; red cells indicate dissimilar ones. In the containment matrices, dark blue ($C = 1.0$) indicates that all trips in the row DayType also appear in the column DayType (full subset relationship). Each DayType is labelled with its trip count and the days of the week on which it is active.}
  \label{fig:matrices-selected}
\end{figure}

Figure~\ref{fig:matrices-selected} presents pairwise Jaccard distance and containment matrices for Otone Kotsu (Japan) and Barrie Transit (Canada). The axes label each DayType with its trip count and active days of week.

\textbf{Otone Kotsu (Japan, 4~DayTypes)} illustrates mixed strategies within one agency. The single-day outlier DT2 is close to the weekend-and-holiday schedule DT0 ($d = 0.16$, $C(\text{DT0}, \text{DT2}) = 0.96$), while the New Year DayType DT1 is almost fully disjoint from all others ($d > 0.94$).

\textbf{Barrie Transit (Canada, 3~DayTypes)} represents separate weekday, Saturday, and Sunday schedules. All pairwise distances are high ($d>0.76$), and no pair shows strong containment ($C_{\max}=0.41$).

\paragraph{Scheduling patterns: similarities and differences}
The two countries share more commonalities than differences. Full nesting ($C = 1.0$ for at least one DayType pair) appears at comparable rates: 69\% of Japanese agencies and 58\% of Canadian agencies exhibit at least one strict subset relationship between DayTypes. Near-identical closest pairs also appear in both countries. Both findings suggest that graduated service reduction---removing trips while keeping the same route structure---is a common practice, not a country-specific one.

The clearest cross-national difference appears in the \emph{extremes} of DayType divergence. Complete disjointness ($d = 1.0$) is more common in Canada (36\%) than in Japan (15\%), and the minimum containment between the most different DayType pairs is substantially lower in Canada on average ($\bar{C}_{\min} = 0.15$ vs.\ $0.35$). This indicates that when Canadian agencies do differentiate their DayTypes, they tend to diverge more sharply---sometimes operating entirely different route sets---while Japanese agencies more consistently maintain overlap across all DayTypes. The maximum pairwise Jaccard distance reflects this: $\bar{d}_{\max} = 0.87$ for Canada vs.\ $0.70$ for Japan.

However, we note that our Canadian sample (38 land-based agencies, predominantly from British Columbia) may not be representative of all Canadian transit, and the Japanese sample (40 feeds) likewise covers a subset of Japan's transit landscape. A larger cross-national study would be needed to confirm whether these patterns reflect genuine national-level scheduling traditions or regional sampling effects.

\section{Discussion}

\subsection{Key Findings}

The central finding is that DayTypes can be treated as identifiable recurring supply classes rather than informal calendar labels. The extraction algorithm identifies meaningful operational patterns from GTFS data across countries, network scales, and calendar-encoding conventions. The case studies recover familiar weekly cycles, holiday service, New Year service, and single-day reduced-service exceptions without external calendar labels.

The second finding is methodological: exact comparison, time-tolerant comparison, and structural-comparability diagnostics answer different questions and should be used together. The exact metric remains valuable because even a one-minute timestamp change may reveal intentional timing adjustment, an operational inconsistency, or a data-production artifact. The time-tolerant metric asks whether the difference persists after bounded same-Route-Pattern timing shifts. The matched-share, unmatched-trip, and trip-count imbalance diagnostics then determine whether it is reasonable to describe the difference as timing variation of a comparable schedule or whether it is primarily a structural service change. This directly addresses the fragility of exact timestamp identity while preserving the planner-facing value of strict schedule audit.

The third finding is empirical: the same exact disjointness statistic can represent different operational mechanisms. In the Canadian sample, many exact-disjoint pairs acquire at least one same-Route-Pattern match under small timing tolerance, but the median matched share remains low and trip-count imbalance remains high. In the Japanese sample, exact-disjoint pairs are even more structurally separated. Thus the proposed metric hierarchy changes the interpretation of the empirical comparison by separating limited timing sensitivity from structurally persistent schedule change.

Extracted DayTypes also provide compact inputs for downstream models. Instead of optimizing or assigning service independently for every date, a model can operate on recurring service classes and represent date-specific deviations around them.

\subsection{Methodological Role}

The contribution of the paper is best understood as a representation layer between raw GTFS data and downstream scheduling or assignment models. Raw GTFS tables encode dates, services, trips, stop times, stops, and shapes separately. Timetable synchronization, vehicle scheduling, deficit-function analysis, network design, and demand assignment models usually require a coherent supply object: a recurrent timetable or service class on which optimization or evaluation is performed. The proposed framework constructs this object explicitly as the defining Route-Pattern-time schedule set of each DayType. The resulting recurring schedule object can serve as an input layer for downstream methodological work, including optimization, assignment, mismatch analysis, and activity-schedule modelling.

This representation has three methodological properties that are important for such downstream use. First, the DayType partition is identifiable from a fixed GTFS feed: for a chosen H3 resolution and valid calendar/trip tables, there is a unique mapping from dates to active trip-specification sets. Second, the Route Pattern keys are invariant to relabeling of administrative GTFS identifiers such as \texttt{stop\_id}, \texttt{shape\_id}, \texttt{trip\_id}, and \texttt{service\_id}, provided the passenger-facing timetable and stop geometry are unchanged. Third, the comparison layer is stable under bounded timing shifts for structurally comparable schedules: exact differences remain visible through $d_0$, while $d_{\epsilon}$, matched-trip share, unmatched-trip count, and trip-count imbalance distinguish timing adjustments from Route Pattern or trip-presence changes.

The resulting metric hierarchy supports a sequence of diagnostic questions. The exact metric asks whether two DayTypes are identical at the trip-specification level. The tolerant metric asks whether apparent differences persist after bounded same-pattern timing shifts. The structural-comparability diagnostics then ask how much of the two schedules is actually matchable and how much service volume differs. This is the sense in which DayType extraction is not only a GTFS grouping operation: it produces identifiable recurring schedule sets and a comparison calculus for timetable optimization, mismatch diagnosis, and schedule-consolidation analysis.

\subsection{Practical Applications}

DayType extraction provides a recurring timetable library for planners and schedulers. Planners can inspect which dates share the same supply pattern, identify nearly identical schedules that may be consolidated, and use the exact--tolerant metric gap together with structural-comparability diagnostics to separate small timing shifts from structural redesigns.

For methodological models, a DayType schedule set can be expanded into the time-space network, deficit-function, vehicle-scheduling, or assignment representation required by a downstream formulation. Demand and activity-schedule models can likewise be stratified by extracted service class, allowing mismatch analysis at the level of recurring operational regimes. Across agencies or feed releases, H3-based Route Pattern keys support comparison without relying on unstable GTFS identifiers.

\subsection{Limitations}

Several limitations remain:

\textbf{Data Quality Dependency:} The algorithm extracts patterns from GTFS data exactly as provided. If the GTFS feed contains errors (missing trips, incorrect times, malformed schedule definitions), these errors propagate into the extracted day types. The exact metric is intentionally sensitive to such issues; the time-tolerant metric helps classify small timing shifts but does not by itself prove whether a difference is intentional or erroneous.

\textbf{Route Pattern Equivalence:} The present framework treats exact H3 stop sequences as Route Pattern identities. This is robust to administrative identifier changes but does not yet model spatially tolerant equivalence between nearly identical stop sequences or shapes.

\textbf{Calendar Interpretation Complexity:} GTFS calendars can encode complex patterns (overlapping service definitions, multiple exception dates, service windows with gaps). While our algorithm handles these correctly, the interpretation of highly complex DayType structures may remain challenging for human operators.

\textbf{Temporal Dynamics:} The current analysis treats each GTFS snapshot independently. Longitudinal extensions are natural because H3 Route Pattern keys are independent of GTFS identifiers: multiple GTFS releases can be aligned, DayTypes can be extracted from each release, and exact/time-tolerant metrics can quantify schedule evolution over time.

\textbf{Demand Integration:} The present paper analyzes scheduled supply. Linking extracted DayTypes to ridership, smart-card records, or activity schedules is required before the framework can diagnose demand-supply mismatch directly.

\subsection{Future Work}

Future work should focus on three extensions. First, longitudinal analysis can track how DayType structures change across GTFS releases and quantify schedule evolution over multiple years. Second, demand integration can relate extracted service classes to ridership, activity schedules, or smart-card data, supporting demand-supply mismatch analysis. Third, spatially tolerant Route Pattern matching can relax exact H3-sequence equality when minor stop or path changes should not imply a structural route redesign.

\section{Conclusion}

This paper establishes Route Patterns and DayTypes as reusable concepts for representing recurring public transit supply. Although these objects are not explicitly defined in GTFS, they are implicitly encoded in calendar and schedule data. The contribution is a methodological representation and comparison layer: it converts raw GTFS tables into identifiable Route-Pattern-time schedule sets and compares those schedules through exact, time-tolerant, and structural-comparability metrics.

We provide formal definitions of Route Patterns and DayTypes, an H3-based method for stable Route Pattern identification, an extraction algorithm for assigning dates to DayTypes, and exact, time-tolerant, and structural-comparability metrics for schedule comparison. The framework has deterministic identifiability for a fixed feed, invariance to administrative GTFS identifier relabeling, and bounded timing-shift stability under the tolerant comparison layer when schedules are structurally comparable. Matched-share, unmatched-trip, and trip-count imbalance diagnostics prevent small same-pattern timing matches from being overinterpreted as whole-schedule timing variation. Validation on Japanese and Canadian feeds shows that standard GTFS data contain compact and interpretable recurring service classes, even when agencies encode calendars differently.

The extracted DayTypes and comparison metrics can serve as compact recurring schedule inputs for timetable synchronization, vehicle scheduling, demand assignment, activity-schedule analysis, mismatch diagnosis, and schedule consolidation.

\section*{CRediT authorship contribution statement}

\textbf{Evgeny Makarov:} Conceptualization, Methodology, Software, Formal analysis, Data curation, Visualization, Writing -- original draft, Writing -- review and editing.
\textbf{Georgy Taubkin:} Conceptualization, Methodology, Writing -- review and editing.

\section*{Declaration of competing interest}

The authors declare that they have no known competing financial interests or personal relationships that could have appeared to influence the work reported in this paper.

\section*{Funding}

This research did not receive any specific grant from funding agencies in the public, commercial, or not-for-profit sectors.

\section*{Data availability}

All GTFS data analysed in this study are publicly available. Japanese GTFS feeds were obtained from the Open Data Portal for Public Transportation (ODPT) at \url{https://ckan.odpt.org}. Canadian GTFS feeds were obtained from the Canadian Public Transit Network Database published by Statistics Canada (Catalogue no.\ 23-26-0003). The extracted DayType sets, pairwise distance and containment matrices supporting the findings of this study are available from the corresponding author upon reasonable request.

\section*{Code availability}

A reference implementation of the DayType extraction and schedule-comparison framework is available as the open-source Python package \texttt{gtfs\allowbreak-daytype}, version~0.1.0 \citep{gtfsdaytype2026}. The package is distributed through \href{https://pypi.org/project/gtfs-daytype/}{PyPI}, and the source code is available on \href{https://github.com/emakarov/gtfs-daytype}{GitHub}. It includes command-line tools, Python APIs, tests, example workflows, and validation examples for the GTFS feeds used in this paper.

\section*{Declaration of generative AI and AI-assisted technologies in the manuscript preparation process}

During the preparation of this work the authors used AI-assisted tools for copy editing to improve readability and style, and to check grammar, spelling and punctuation. AI tools were also used to automate portions of the data processing pipeline; all AI-assisted outputs were manually reviewed by the authors and cross-referenced against results obtained by independent calculation methods to verify correctness. After using these tools, the authors reviewed and edited the content as needed and take full responsibility for the content of the published article.

\end{document}